\documentclass{pasj01}

\usepackage{graphicx}
\usepackage{bm}
\usepackage{natbib}
\usepackage{mathrsfs}
\usepackage{comment}

\Received{$\langle$reception date$\rangle$}
\Accepted{$\langle$acceptance date$\rangle$}
\Published{$\langle$publication date$\rangle$}

\begin{document}

\title{Constraining the luminosity function of active galactic nuclei through the reionization observations in the SKA era}


\author{Kai T.\ \textsc{Kono},\altaffilmark{1}$^{*}$
 Taichi \textsc{Takeuchi},\altaffilmark{1}
 Hiroyuki \textsc{Tashiro},\altaffilmark{1}$^{*}$ 
 Kiyotomo \textsc{Ichiki},\altaffilmark{1,2}$^{* \dag}$ 
 Tsutomu T.\ \textsc{Takeuchi},\altaffilmark{1,3}$^{*}$
 }
\altaffiltext{1}{Division of Particle and Astrophysical Science, Nagoya University, Nagoya 464--8602, Japan}
\altaffiltext{2}{Kobayashi-Maskawa Institute for the Origin of Particles and the Universe, Nagoya University, Nagoya 464--8602, Japan}
\altaffiltext{3}{The Research Center for Statistical Machine Learning, the Institute of Statistical Mathematics, 10--3 Midori-cho, Tachikawa, Tokyo 190--8562, Japan}
\email{kono.kai@c.mbox.nagoya-u.ac.jp, hiroyuki.tashiro@nagoya-u.jp, ichiki.kiyotomo@c.mbox.nagoya-u.ac.jp, takeuchi.tsutomu@g.mbox.nagoya-u.ac.jp}


\maketitle

\begin{abstract}
Ultraviolet (UV) and X-ray photons from active galactic nuclei (AGNs) can ionize hydrogen in the intergalactic medium (IGM). 
We solve radiative transfer around AGNs in high redshift to evaluate the $21\mbox{-}{\rm cm}$ line emission from the neutral hydrogen in the IGM and obtain the radial profile of the brightness temperature in the epoch of reionization. 
The ionization profile extends over $10\; [{\rm Mpc}]$ comoving distance which can be observed in the order of $10\; {[\rm arcmin]}$.
From estimation of the radio galaxy number counts with high sensitivity observation through the Square Kilometre Array (SKA), we investigate the capability of parameter constrains for AGN luminosity function with Fisher analysis for three evolution model through cosmic time. 
We find that the errors for each parameter are restricted to a few percent when AGNs are sufficiently bright at high redshifts. 
We also investigate the possibility of further parameter constraints with future observation beyond the era of SKA.
\end{abstract}

\section{Introduction}
\label{sec:Intro}

Supermassive black holes (SMBH) are observed in the center of almost all massive galaxies \citep{2005SSRv..116..523F}.
Their mass is often $> 10^9~{\rm M}_\odot$.
Many observations suggest that SMBH masses are related to the masses and/or velocity dispersions in the bulges of host galaxies \citep[e.g.,][]{1998AJ....115.2285M,2000ApJ...539L...9F,2000ApJ...539L..13G,2003ApJ...589L..21M}.
This fact implies the co-evolution of an SMBH and a host galaxy.
There are many works about the role of SMBHs in the galaxy
evolution \citep[see, e.g., ][and references therein]{2009Natur.460..213C}. 
However, the origin of SMBHs, when and how SMBHs formed, is one of the biggest challenges in the structure formation of the Universe.
High-redshift quasar (QSO) surveys revealed that SMBHs already existed at~$z>6$ \citep{2006AJ....131.1203F,2011Natur.474..616M,2015Natur.518..512W}. 
To explain the formation of the SMBHs in such high redshifts, many scenarios are proposed.
Although it is widely accepted that SMBHs have evolved from a kind of seeds,
various SMBH seeds are claimed, including the remnants of first stars \citep{2001ApJ...551L..27M,2009ApJ...701L.133A,2012ApJ...756L..19W}, the direct collapse of massive gas cloud \citep{1994ApJ...432...52L,1995ApJ...443...11E,2006MNRAS.370..289B}, and the primordial black holes formed in the very early universe \citep{2002PhRvD..66f3505B,2004PhRvD..70f4015D}.
Observations of SBMHs at high-redshifts are strongly desired to restrict the seed scenarios. 

One of the observables to explore SMBHs in the early Universe is active galactic nuclei (AGNs). 
X-ray surveys are efficient to investigate the evolution and population of AGNs at high redshifts.
Wide and deep X-ray surveys by XMM and Chandra provided large samples of AGNs, 
and consequently the abundance of AGNs in $z \sim 1\mbox{--}5$ is well studied \citep{2015A&ARv..23....1B}. 
Nonetheless, the information on AGNs in $z>5$ is still limited. 
Currently, about 30 sources are found in $z>6.5$ \citep{2020MNRAS.491.3884P,2018ApJ...856L..25B, 2018A&A...614A.121N}.
Recently it is claimed that a significant population of faint AGNs would exist at $4< z < 6.5$ \citep{2015A&A...578A..83G}. 
The existence of such faint AGNs would give a strong impact on 
the ionization process during the epoch of reionization~(EoR)~\citep{2015ApJ...813L...8M,2017MNRAS.471.3713Y,2018MNRAS.473.1416M}. 
 
The EoR is an era at which the intergalactic medium~(IGM) has drastically changed from neutral to highly ionized.
Cosmological observations, including CMB, distant QSO and galaxy observations
revealed that the cosmic reionization process has completed before $z\sim6$.  
Cosmic reionization is driven by ionizing photons with an energy of $E>13.6\ {\rm eV}$. 
The UV radiation from newly formed massive stars is thought to be the main source of the ionizing photons. 
However, AGNs are still a candidate of a significant contributor to the reionization. How much star forming galaxies and AGNs contribute to the cosmic reionization has still been actively debated \citep{2012MNRAS.425.1413F, 2012ApJ...752L...5B}. 
Further, since the typical UV slopes of star forming galaxies and AGN are different, the radial ionization profile around these sources are very different. 
Thus, it is necessary to determine the shape of luminosity function of ionizing sources at high redshifts for the determination of ionizing history of the Universe. 
Some literature are trying to quantify the evolution of luminosity function in the EoR \citep{2015A&A...578A..83G, 2015ApJ...813L...8M, 2018MNRAS.474.2904P} to quantify cosmic reionization.



Redshifted 21-cm line observations can be expected to provide new constraints on the AGN luminosity function in high redshifts.
Measurement of the 21-cm radiation from neutral hydrogen in the IGM is useful as a probe of the physical properties of the IGM reionization process in high redsfhits \citep[for a review, see][]{2006PhR...433..181F}.
Currently, the detection of 21~cm signals from the epoch of reionization has not been confirmed yet. 
The Square Kilometre Array (SKA) project is expected to measure the 21-cm signals from the epoch of reionization and even from the Cosmic Dawn \citep{2015aska.confE...1K}.
Since AGNs emit ionizing photons and ionize the IGM, AGNs can contribute to generating the spatial fluctuations of 21$\mbox{-}$cm signals \citep{2017MNRAS.469.4283K}.
In addition, AGNs can serve as X-ray sources.
Since X-ray heats the IGM and produce Lyman-$\alpha$ photons which can excite the hyperfine structure of neutral hydrogen, it would also affect the 21-cm signals \citep{2007MNRAS.376.1680P,2013JCAP...09..014C,2014Natur.506..197F}.
The constraint on the number of X-ray sources obtained from redshifted 21$\mbox{-}$cm measurements,then, can provide useful information about the AGN population.


In this paper, we investigate the feasibility of the AGN number count by using redshifted 21$\mbox{-}$cm observations in the SKA era.
Luminous objects before the epoch of reionization can make distinctive signal structures on redshifted 21$\mbox{-}$cm signal maps.
Some preceding papers have been published to investigate the signal profiles of first stars, galaxies, quasars and primordial black holes \citep{2007MNRAS.375.1269Z,2008ApJ...684...18C,2013MNRAS.435.3001T,2014MNRAS.445.3674Y}.
Therefore, we can expect to find luminous objects by 21$\mbox{-}$cm surveys. 
Focusing on the 21$\mbox{-}$cm signal created by individual AGNs, we investigate the number count of AGNs with a simple analytic model of the luminosity function of AGNs.
Then, using the number count dependence on the redshift and angular resolution,  we demonstrate how well we can recover the luminosity function through the SKA observation.

This paper is organized as follows. In Section~\ref{sec:AGNLF}, we describe AGN luminosity function and galaxy evolution models we assumed. 
Section~\ref{sec:21cm} is dedicated to calculate the 21$\mbox{-}$cm signal around an AGN to estimate the limit luminosity with the SKA sensitivity. 
In Section~\ref{sec:AGNNC}, we presents our results in galaxy number counts and the capability of galaxy evolution parameter constraints and discuss for future observations. 
Finally, we conclude in Section~\ref{sec:Summary}.

\section{AGN luminosity function (LF)}
\label{sec:AGNLF}

AGN activities in high redshift~($z>8$) could provide huge impact on cosmological structure formation history, as we discussed in Section~\ref{sec:Intro}.
However, either theoretically or observationally, the AGN luminosity function (LF) at high redshifts~($z>8$) is totally uncertain .
The aim of this paper is to investigate the feasibility of future 21$\mbox{-}$cm observation including SKA for probing the AGN LF.
To demonstrate this, we introduce a simple model of the AGN LF.
\begin{eqnarray}
   \frac{\mathrm{d}n}{\mathrm{d}\log{L}}=(1+z)^{Q(z)}\frac{A}{\displaystyle \left(\frac{L}{L_*}\right)^{\gamma_1}+\left(\frac{L}{L_*}\right)^{\gamma_2}},
\end{eqnarray}
where $L_*$ is the critical luminosity of the power-law index,
$\gamma_1$ and $\gamma_2$~($\gamma _1 < \gamma_2$)~are the power-law indices for the lower and higher luminosity sides, respectively, $A$ is a normalizing factor at the normalized luminosity $L_{\rm n} = 10^{10}~[L_\odot]$, and $L_*$ is the characteristic luminosity that determines transition of the power-law indices.
In the model, $Q(z) $ represents the redshift evolution of the LF.
This type of formulation has been often used for the studies on the far-infrared (FIR) galaxy evolution \citep[e.g.,][]{1990ApJ...358...60L,1990MNRAS.242..318S,1999MNRAS.308..897L}, and recalling the tight relation between X-ray and FIR emission, the assumption is naturally justified \citep[e.g.,][]{1992ApJ...388...82D}.  
We take the first order of polynomial of redshift as
\begin{eqnarray}
   Q(z)=\beta_1+\beta_2 z \; .\label{eq:ev_model}
\end{eqnarray}
where the parameters $\beta_i$ are the power-law indices for the redshift dependence.
Therefore, in our model, the model parameter set is 
${\bf \theta} = \{A, L_{10*}, \gamma_1, \gamma_2, \beta_1, \beta_2\} $, where $L_{10*}$ is defined as $L_{10*} = L_*/L_{\rm n}$.
This double power-law~(DP) model agrees well with the AGN observation at low redshifts, and commonly used as a proper description of the AGN LF (e.g., \citealt{2007MNRAS.375..931M}; \citealt{2014ApJ...786..104U}).
Therefore, we adopt this model even for a high redshift LF which we are interested in this paper. 

\begin{table}
 \begin{center}
 \begin{tabular}{l|lccccc} \hline
  & $\mathscr{A}$ & $\gamma_1$ & $\gamma_2$ & $\beta_1$ & $\beta_2$ & $L_{10*}$\\ \hline
  Model~I& $10^{-4}$& 1.5 & 3.5 & $-1.0$ & $-0.02$ & 0.85 \\
  Model~II& $10^{-4.5}$& 1.5 & 3.5 & 1.0 & $-0.3$ &  0.85 \\ 
  Model~III& $10^{-5}$& 1.5 & 3.5 & 2.0 & $-0.5$ & 0.85 \\ \hline
 \end{tabular}
 \end{center}
 \caption{Parameter set for our DP models of the AGN LF in high redshifts.}
 \label{tab:paraD}
 \end{table}

In order to choose the fiducial parameter set of our model, it is useful to calculate the UV emissivity and compare it with high redshift AGN observation.
The emissivity at the energy $E_{912}$ corresponding to the wavelength, $912$~\AA, can be obtained by
\begin{eqnarray}
  \epsilon_{912} = \int L_E (L;E_{912}) \frac{{\rm d}n}{{\rm d}L}{\rm d} L \;,
\end{eqnarray}
where $L_E (L;E) $ is the energy spectrum of AGNs with the total luminosity $L$.
For simplicity, we assume that AGNs have a power-law energy spectrum 
with the spectral index $p$,
\begin{eqnarray}
 L_E (L;E) = \mathscr{A}(L) \left(\frac{E}{E_{912}} \right)^{p}
 \quad(E_{912} < E < 100~{\rm keV})\; , \label{eq:power-law}
\end{eqnarray}
where we set $p=-1$ as suggested by \citet{2005MNRAS.363.1069K},
$E_{\rm min} = 200$~eV and $\mathscr{A}$ is the normalization factor which satisfies
the relation~$L = \int L_E(L;E){\rm d}E$ over the integration interval $E_{912}
 z < E < 100$~keV.

\begin{figure}
\begin{center}
  \includegraphics[width=0.5\textwidth]{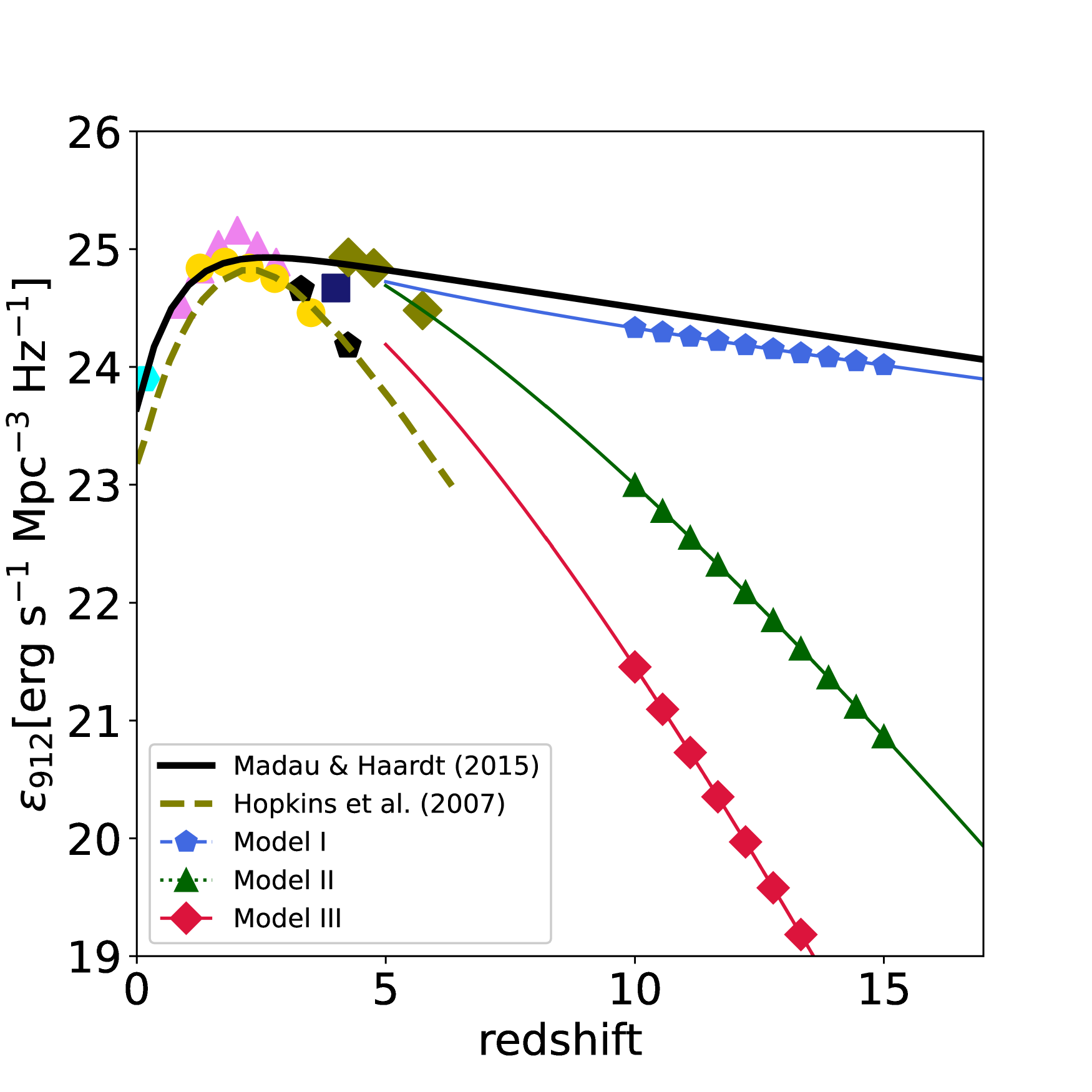}
 \end{center}
 \caption{The expected emissivity of ionizing photon from AGN during cosmic time (pentagons, triangles, diamonds, and solid lines) for our DP models. The black solid line and olive dashed line are empirical model of emissivity obtained by \citet{2015ApJ...813L...8M} and \citet{2007ApJ...654..731H}, respectively. 
Observations are presented for comparison as follows; 
cyan pentagon: \citet{2009A&A...507..781S}, 
pink triangles: \citet{2013A&A...551A..29P}, 
yellow circles: \citet{2007A&A...472..443B},
black pentagons: \citet{2012ApJ...755..169M}, 
dark blue square: \citep{2011ApJ...728L..26G}, 
olive diamonds: \citet{2015A&A...578A..83G}, respectively.}
\label{fig:e912}
\end{figure}

Figure~\ref{fig:e912} 
presents the emissivity, $\epsilon_{912}$, for our three fiducial parameter sets.
We summarize our parameter set in table~\ref{tab:paraD}.
Currently $\epsilon_{912}$ is measured in the redshifts lower than $z=7$.
We plot the several observation data
in colored marks.
The figure tells us that our fiducial models lie
inside the scatter of the observation data in lower redshifts.
For comparison, we plot two empirical models proposed by \cite{2015ApJ...813L...8M} and \cite{2007ApJ...654..731H} in the black and olive solid lines, respectively.

In our fiducial models, Model~I has the strongest emissivity. 
In this model, the redshift evolution is similar to the model of \cite{2015ApJ...813L...8M}, but they constructed it based on optically selected AGN samples.
In Model~III, we choose the parameters to match the redshift evolution of $\epsilon_{912}$ to the galaxy evolution model at low redshifts \citep{2007ApJ...654..731H} 
which is based on multiple observations on X-ray, near- and mid-IR bands.

\section{21-cm signal around an AGN}
\label{sec:21cm}

The AGNs can ionize and heat the surrounding IGM. 
As a result, it is expected that the unique spatial distribution of 21~cm brightness temperature is observed around AGNs as the 21-cm signals of AGNs.
The strength and size of the signal depends on the AGN luminosity.
Therefore, we can count the abundance of AGNs with $L$ through measuring the 21-cm signals by future 21-cm observation.
In this section we calculate the 21-cm signal produced by AGNs with $L$.

\subsection{IGM neutral fraction and temperature around an AGN}

The evaluation of the 21-cm signal requires the spatial distribution of the neutral fraction~$x_{\rm HI}$ and gas temperature~$T_{\rm k}$ of hydrogen around an AGN.
To obtain these distributions, we follow the method in \citet{2007MNRAS.375.1269Z}.

As mentioned in the previous section, we assume that AGNs have the power-law energy spectrum, Eq.~(\ref{eq:power-law}).
With the assumption of the isotropic photon emission, the energy flux at distance~$r$ from an AGN
can be obtained by
\begin{eqnarray}
\mathscr{F}(E,r) =e^{-\tau(E,r)} \frac{L_E(E)}{4\pi r^2}.
\label{eq:e-flux}
\end{eqnarray}
Here 
$\tau(E,r)$ is the optical depth for
a photon with energy~$E$ from the AGN to the distance~$r$,
\begin{eqnarray}
  \tau(E,r) = \int^r_0 n_{\rm H} x_{\rm HI}(r) \sigma(E) {\rm d}r,
\end{eqnarray}
where $n_{\rm H}$ is the IGM hydrogen number density and
$x_{\rm HI}(r)$ is the hydrogen neutral fraction at~$r$.
Taking into account the helium contribution,
the cross-section~$\sigma(E)$ is given by
\begin{eqnarray}
\sigma(E) = 
\sigma_{\rm H}(E) + \frac{n_{\rm He}}{n_{\rm H}} \sigma_{\rm He}(E),
\end{eqnarray}
where $\sigma_{\rm H}(E) = \sigma_0 (E_0/E)^3$ with
$\sigma_0 = 6 \times 10^{-18}~{\rm cm^2}$
and $E_0 = 13.6 ~{\rm eV}$.
For the helium contribution,
we set the number ratio of helium to hydrogen
to $n_{\rm He}/n_{\rm H} = 1/12$ and
take $\sigma_{\rm He}(E)$ as \cite{1994MNRAS.269..563F}
\begin{eqnarray}
  \sigma_{\rm He}(E) = 1.13 \times 10^{-14}\left( \frac{1}{E^{2.05}} - \frac{9.775}{E^{3.05}} \right)~[{\rm cm^2}] \; .
\end{eqnarray}

Next we consider the IGM ionization by an AGN with the energy flux given by Eq.~(\ref{eq:e-flux}). 
In the calculation of the ionization,
we simply assume that the photoionization due to the AGN
is balanced by recombination. Therefore, the neutral fraction of
hydrogen~$x_{\rm HI}$ is obtained by
solving the equilibrium equations between them \cite{2005MNRAS.360L..64Z},
\begin{eqnarray}
\alpha^{({\rm B})} n_{\rm H}^2 (1-x_{\rm HI})^2 
= \Gamma(r) n_{\rm H} x_{\rm HI} 
\left( 1+ \frac{\sigma_{\rm He}}{\sigma_{\rm H}} \frac{n_{\rm He}}{n_{\rm H}} \right),
\label{eq:ionization}
\end{eqnarray}
where $\alpha^{(\rm B)}$ is the case-B recombination 
rate,~$\alpha^{({\rm B})} = 2.6\times 10^{-13} T_4^{-0.85} {\rm cm^3 \ s^{-1}}$
with $T_4 = T_{\rm k}/10^4~[{\rm K}]$.
In Eq.~(\ref{eq:ionization}),
$\Gamma(r)$ represents the photoionization rate per a hydrogen atom at the distance~$r$ from the AGN,
\begin{eqnarray}
  \Gamma(r) = \int^{\infty}_{E_0} \sigma(E) \mathscr{F}(E,r) 
\left[ 1+ \frac{E}{E_0} \phi (E,x_{\rm e}) \right] \frac{{\rm d}E}{E}.
\end{eqnarray}
Here we introduce the function $\phi(E,x_{\rm e})$, which provides the fraction of the energy used for the secondary ionizations over the injected energy from an AGN with $x_{\rm e} =1-x_{\rm HI}$.
For $\phi(E,x_{\rm e})$, we adopt the fitting formula in \citet{1985ApJ...298..268S} and \citet{2004ApJ...613..646D},
\begin{eqnarray}
 \phi(E,x_{\rm e}) &=& 0.39 \Big[ 1- x_{\rm e}^{0.4092 \ a(E,x_{\rm e})} \Big]^{1.7592}, \nonumber \\
 a(E,x_{\rm e}) &=& \frac{2}{\pi} \arctan \left[ 
 \left( \frac{E}{0.12{\rm keV}} \right)
 \left( \frac{0.03}{x_{\rm e}^{1.5}} + 1 \right)^{0.25} \right].
\end{eqnarray}

\begin{figure}
\begin{center}

\includegraphics[width=0.4\textwidth]{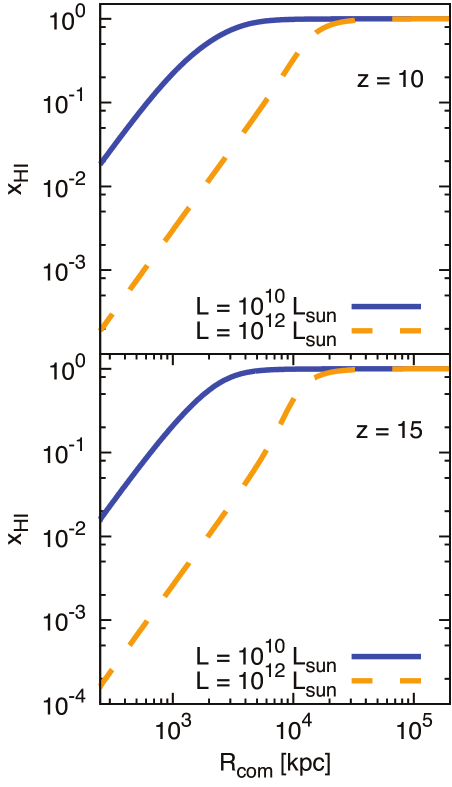}
\end{center}
 \caption{The radial profile of neutral fraction around an AGN.
 The blue and orange lines are for $L=10^{10}~L_\odot$ and $L=10^{12}~L_\odot$, respectively. We set $z=10$ and $12$ in the top and bottom panels, respectively. }
 \label{fig:xhi}
\end{figure}

We plot the results as the radial profiles of the neutral fraction in
Figure~\ref{fig:xhi}.
In the figure, we set the AGN luminosity to 
$L= 10^{10}~\rm L_\odot$ and $10^{12}~\rm L_\odot$.
As the luminosity increases, the ionized region becomes large. When the
AGN has the Eddington luminosity with $10^7~\rm M_\odot$, the ionized
region expands to $10$~Mpc scales.
Figure~\ref{fig:xhi} also shows us the redshift dependence of the ionized region.
Since the ionization and recombination process 
depends on the number density, the ionized region is smaller 
in higher redshift in physical scales. Note that 
Figure~\ref{fig:xhi} is represented in the comoving scale, and therefore the redshift dependence is not obvious.

The X-ray photons emitted from the AGN can also heat the IGM.
To obtain the temperature heated by the AGN, we take two assumptions.
The first assumption is that the heating rate is constant during the AGN
lifetime~$t_{\rm life}$. The other is that the cooling effects are negligible.
In the IGM, the main cooling mechanisms are the expansion of the
Universe and the Compton cooing with CMB photons.
Compared with the AGN lifetime, which we take~$t_{\rm life} \sim 10$~Myrs,
these time scales are longer at redshifts $10<z< 20$.

With these assumptions, the temperature at the distance~$r$, $T_{\rm k}(r)$,~is
given by
\begin{eqnarray}
  T_{\rm k}(r) = \frac{2}{3} \frac{\mu \mathscr{H}(r) t_{\rm life} }{ n_{\rm H} k_{\rm B}} \; , \label{eq:gas_temp}
\end{eqnarray}
where $\mu$ is the mean molecular weight and $k_{\rm B}$ is the Boltzmann constant.
In Eq.~(\ref{eq:gas_temp}), $\mathscr{H}(r)$ is the heating rate per unit volume at distance~$r$ which can be written as
\begin{eqnarray}
  \mathscr{H}(r) = f n_{\rm H} x_{\rm HI}(r) \int^{\infty}_{E_0} \sigma(E) \mathscr{F}(E,r) {\rm d}E \; , \label{eq:heating}
\end{eqnarray}
where $f$ represents the fraction of the photon energy which is transferred to the IGM temperature through the collisional excitation of the IGM.
We used the fitting formula for~$f$ provided by \citet{1985ApJ...298..268S};
\begin{eqnarray}
  f = C[1-(1-x_{\rm e}^a)^b]
\end{eqnarray}
with $C=0.9771,\, a=0.2663,\, b=1.3163$, imposing the lower limit as $f=0.11$ for $x<10^{-4}$ as modified in \citet{2007MNRAS.375.1269Z}.
As the distance~$r$ increases, the heating efficiency becomes low.
At a sufficient distance, the temperature should correspond to the background baryon temperature~$\overline{T}_k(z)$.
Therefore, if $T_{\rm k}(r) $ in Eq.~(\ref{eq:gas_temp}) becomes lower than the background baryon temperature at $r_{*}$, we set $T_{\rm k}(r) = \overline{T}_k(z)$ at $r> r_*$.

\begin{figure}
\begin{center}
\includegraphics[width=0.4\textwidth]{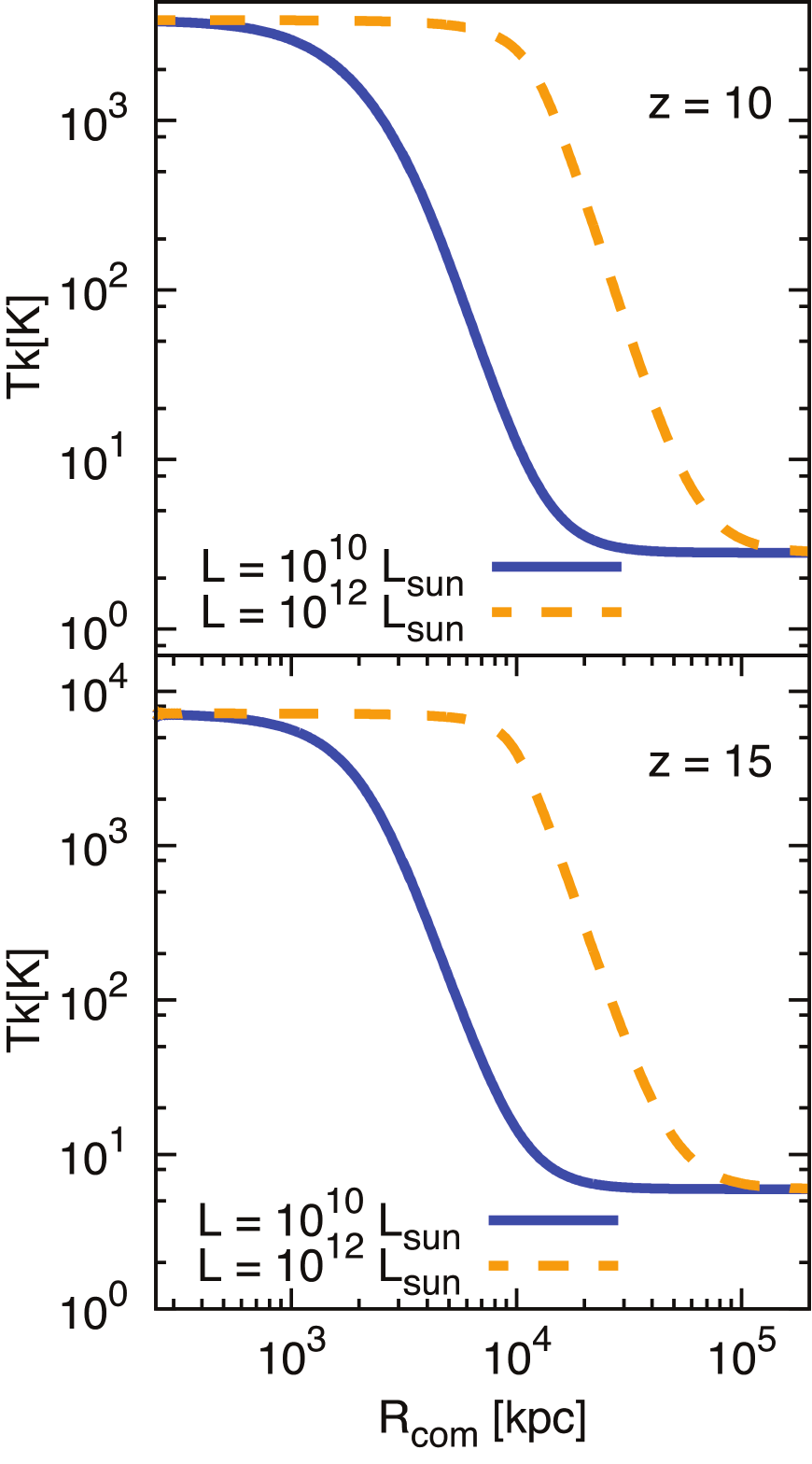}
\end{center}
 \caption{The radial profile of kinetic temperature around an AGN.
 The colored lines are same as in Figure~\ref{fig:xhi}.}
\label{fig:tk}
\end{figure}

We plot the kinetic temperature radial profile around an AGN in
Figure~\ref{fig:tk}.
Similarly to the case of the ionization, as the luminosity increases, the AGN can heat up neutral hydrogen in
the IGM at further distance.
In particular, AGNs with $L=10^{12}~L_\odot$ can heat the IGM even at comoving $100~$Mpc distance.
In the inner side which is highly ionized, the temperature does not depends on the distance and the AGN luminosity.
This is because the heating rate, Eq.~(\ref{eq:heating}), is almost
constant with the equilibrium assumption in~Eq.~(\ref{eq:ionization}) in the highly ionized
region,~$x_{\rm HI} \ll 1$. In other words, once the region is highly ionized, there happens no
additional heating any more and the temperature gets saturated. 
Figure~\ref{fig:tk} also tells us the redshift dependence of the temperature.
The heating rate grows as the redshift increases. Therefore, the resultant
heated temperature is also large in high redshifts.

\subsection{Different brightness temperature}


In observations of redshifted 21-cm signals, the observed value is described as the difference of the brightness temperature at the observed frequencies from the CMB temperature.
This different brightness temperature is given by \citep{1997ApJ...475..429M,2003ApJ...596....1C}
\begin{eqnarray}
  \delta T_{\rm{b}}\equiv T_{\rm{b}}-T_{\rm{CMB}}(z)=\frac{[T_{\rm s}-T_{\rm{CMB}}(z)](1-e^{-	\tau})}{1+z} \; , \label{deltatb}
\end{eqnarray}
where~$T_{\rm CMB}$, $T_{\rm{b}}$ and ${T_{\rm s}}$
represent the CMB temperature, the 21-cm brightness temperature
and its spin temperature, respectively. 
In the equation,~$\tau$ is 
the 21-cm optical depth~of the IGM,
\begin{eqnarray}
  \tau=\frac{3}{4}\frac{h_{\rm{P}} c^{3} A_{10}}{8\pi \nu_{10}^{2}  k_{\rm{B}}}\frac{x_{\rm HI}n_{\rm H}}{T_{\rm s}H(z)} \; , \label{eq:tau1}
\end{eqnarray} 
where
$A_{10}$ is the
Einstein A-coefficient for the transition, 
$\nu_{10}$ is the frequency corresponding to the energy difference
between the transition states, $h_{\rm{P}}$ is the Planck constant and $c$ is the speed of light.
In Eq.~(\ref{deltatb}),
we ignore the effects of 
the peculiar velocity and the thermal velocity of the gas
on the velocity gradient along the line-of-sight, which are 
generally smaller than the Hubble expansion effect considered in
Eq.~(\ref{eq:tau1}).

Since we are interested in the 21-cm signal from the IGM, the optically thin limit, $\tau \ll 1$, is valid. 
Therefore, we can approximate the different brightness temperature as
\begin{eqnarray}
 \delta T_{\rm b}\simeq
 \frac{3}{32\pi}\frac{h_{\rm{P}}c^{3}A_{10}}{k_{\rm{B}}\nu_{10}^{2}}
 \frac{x_{\rm HI} n_{\rm{H}}}{(1+z)H(z)} \left[ 1-\frac{T_{\rm{CMB}}(z)}{T_{\rm{s}}} \right] \; .  \label{eq:dtb2}
\end{eqnarray}

The spin temperature of the 21-cm lines in the cosmological context
is given by \citep{1952AJ.....57R..31W,1958PIRE...46..240F}
\begin{eqnarray}
  T_{\rm s} = \frac{T_{\rm CMB} + y_{\rm k} T_{\rm k} + y_{\alpha} T_{\rm k}} {1 + y_{\rm k} + y_{\alpha}} \; , \end{eqnarray}
where $y_{\rm k} $ and $ y_{\alpha}$ are the kinetic coupling and Ly${\alpha}$ coupling coefficients, respectively.

The contributions to $y_{\rm k}$
are divided into three collision terms of neutral hydrogen
with neutral hydrogen, electrons and protons as
\begin{eqnarray}
  y_{\rm k} = \frac{T_{\ast}}{A_{10} T_{\rm k}} (C_{\rm H} + C_{\rm e} + C_{\rm p}),
\end{eqnarray} 
where 
$C_{\rm H}$, $C_{\rm e}$ and $C_{\rm p}$
are the collisional de-excitation rates due to neutral hydrogen, electrons and protons and $T_*$ is the temperature corresponds to the energy difference between singlet and triplet of electron, namely $T_*=0.0681\ [{\rm K}]$.
For these rates, we adopt the fitting formulae given by \citet{2006ApJ...637L...1K},
\begin{eqnarray}
 &&C_{\rm H} = 3.1 \times 10^{-11} n_{\rm H} T_{\rm k}^{0.357} 
		\exp{ \left( - \frac{32}{T_{\rm k}} \right) }, \\
 &&C_{\rm e} = n_{\rm e} \gamma_{\rm e} ,\\
 &&C_{\rm p} = 3.2{C_{\rm H}}\frac{n_{\rm p}}{n_{\rm H}},
\end{eqnarray}
where $\gamma_e$ represents the de-excitation coefficient for ${\rm e\mbox{-}H}$ collision, which is fitted with the temperature $T_{\rm k}$ by
\begin{eqnarray}
 &&\log_{10}\left(\frac{\gamma_{\rm e}}{1 ~[{\rm cm^3 s^{-1}}]}\right) \nonumber \\ 
 && \qquad= -9.067 + 0.5 \log_{10}T_{\rm k} 
		\exp{ \left[ - \frac{(\log_{10}T_{\rm k})^{4.5}}{1800} \right] }.
\end{eqnarray}

The Lyman-$\alpha$ coupling coefficient is provided in \citet{1958PIRE...46..240F}
\begin{eqnarray}
  y_{\alpha} = \frac{16 \pi^2 T_{\ast} e^2 f_{12} J_0(r)}{27 A_{10} T_{\rm k} m_{\rm
e} c} \; , \label{eq:y-alpha}
\end{eqnarray}
where $f_{12}$ is the oscillator strength of the Lyman-$\alpha$ transition, $f_{12}=0.416$, and $J_0(r)$ represents the Lyman-$\alpha$ intensity at distance $r$ from the AGN. 
{}To obtain $J_0(r)$, we consider the secondary collisional excitation by electrons released in the photoionization by the AGN.
In this case, the Lyman-$\alpha$ intensity can be written as \citep{2007MNRAS.375.1269Z}
\begin{eqnarray}
  J_0(r) = \frac{\phi_{\alpha}c}{4 \pi H(z) \nu_{\alpha}} n_{\rm H} x_{\rm HI}
(r) \int^{\infty}_{E_0} \sigma(E) \mathscr{F}(E,r) {\rm d}E,
\end{eqnarray}
where $\nu_\alpha$ is the Lyman-$\alpha$ frequency and $\phi_\alpha$ is the energy fraction for the secondary excitation.
We set $\phi_{\alpha} = 0.48 \times (1-x_{\rm e}^{0.27})^{1.52}$, according to \citet{1985ApJ...298..268S}.

\begin{figure}
\begin{center}
\includegraphics[width=0.4\textwidth]{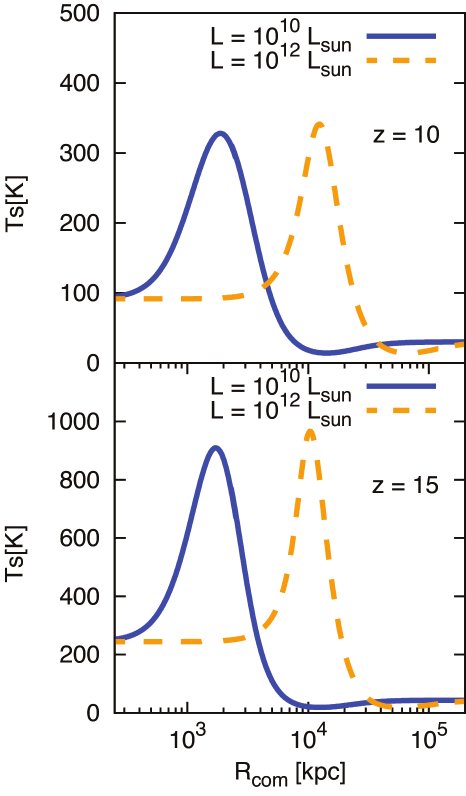}
\end{center}
\caption{The radial profile of spin temperature $T_{\rm s}$ around an AGN.
 The colored lines are same as in Figure~\ref{fig:xhi}.}
\label{fig:tspin}
\end{figure}

Figure~\ref{fig:tspin} represents the spin temperature profile around an AGN.
As shown in Eq.~(\ref{eq:y-alpha}), the Lyman-$\alpha$ coupling is proportional to $x_{\rm HI}$. 
Therefore, in the ionized~region, $y_\alpha$ is not effective.
As a result, the spin temperature is controlled by $y_{\rm k}$.
Since the kinetic temperature is constant in the ionized region, the resultant spin temperature also keeps constant.
As $x_{\rm HI}$ increases, $y_\alpha$ becomes larger than $y_{\rm k}$ and has an important role to determine the spin temperature. The Lyman-$\alpha$ intensity is stronger where $x_{\rm HI}$ is close to a unity, and the spin temperature reaches the
maximum value. 
Moreover, the maximum value of the spin temperature is not sensitive to the AGN luminosity.
This fact means that $J_0$ at the peak position of the spin temperature is independent of the AGN luminosity.
This reason is same as in the discussion about the independence of the maximum gas temperature on the AGN luminosity.
The Lyman-$\alpha$ photon production has been saturated at the peak position and, as a result, does not depend on the AGN luminosity.
For the further discussion we refer the reader to  \citet{2007MNRAS.375.1269Z}.

Even at the cosmological distance from the AGN, the Lyman-$\alpha$ coupling can deviate the spin temperature from the CMB temperature.
When the AGN luminosity is enough large as the Eddington luminosity with $L > 10^{12} ~\rm L_\odot$, the spin temperature is well deviate even at the comoving distance larger than 10~Mpc.
Since the IGM is cooler than the CMB, the spin temperature is lower than the CMB temperature in this range.
At a larger distance, the spin temperature finally approaches the CMB temperature.

In Figure~\ref{fig:tspin}, we also show the redshift dependence of the spin
temperature.
Since the coupling coefficients, $y_{\rm k}$ and $y_\alpha$, are
proportional to the IGM hydrogen density $n_{\rm H}$,
they becomes large when the redshift increases.
Additionally, the gas temperature is also large in high redshifts.
Therefore, the resultant spin temperature is larger in high redshifts.
The maximum value of the spin temperature at $z=15$ is almost two times
higher than at $z=10$.

\begin{figure}
\begin{center}
\includegraphics[width=0.4\textwidth]{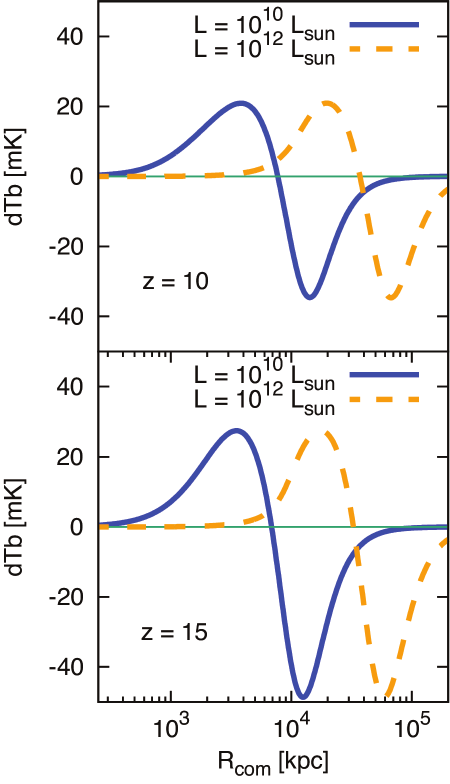}
\end{center}
\caption{The radial profile of brightness temperature $\delta T_{\rm b}$ around an AGN. The colored lines are same as in Figure~\ref{fig:xhi}.}
\label{fig:dtb}
\end{figure}

Based on the spin temperature profiles in~Figure~\ref{fig:tspin}, 
we can calculate the differential brightness temperature
around the AGN through Eq.~(\ref{deltatb}) and~(\ref{eq:tau1}).
We plot the results in Figure~\ref{fig:dtb}.
The region close to the AGN is ionized and the 21-cm signal
vanishes there.
Then gradually the positive~(emission) signal arises where
the gas temperature is much higher than the CMB temperature.
At the sufficient distance, the signal becomes negative~(absorption)
and reaches the negative peak.

As shown in Eq.~(\ref{eq:dtb2}), when the spin temperature is
larger than the CMB temperature~(i.e., the emission signal case),
the signal amplitude is saturated. However, the smaller the spin
temperature, the larger the amplitude of the absorption signal is.
Additionally, the absorption signal region has larger volume than the
emission signal region. Therefore, the detection of the absorption
signal can be easier than the detection of the emission signal. 

The peak amplitude in both negative and positive sides is almost independent of
the AGN luminosity,
because of the same reason as in the cases of the
gas temperature and spin temperature.
On the contrary, the peak position depends on the AGN luminosity.
Therefore, measuring the distance of the negative peak from the AGN~(or the size
of the negative signal region) provides with an important information
to know the luminosity of the AGN at the center of the signal region.

\section{AGN number count}
\label{sec:AGNNC}

Now we present whether future 21-cm observation, such as SKA, can access the AGN LF in high redshifts to demonstrate the AGN number count with our three fiducial models.
First, we evaluate the detectability of the 21-cm AGN signals obtained in the previous section with considering the current design of the SKA observation.
Then to investigate the possible constraint on the LF, we conduct a Fisher matrix analysis for our fiducial models.

\subsection{Detectability of the AGN signals by SKA}
\label{Sec:detectability}

To evaluate the signal detected by SKA, we need to take into account the finite
resolutions of the observation in the frequency~(the line-of-sight direction) and the angular direction (the direction perpendicular to the line-of-sight direction).
We can obtain the predicted signal for the observation by smoothing the signal profile with these resolutions.

First, we make the 3-dimensional signal map from the radial profile of the different brightness temperature, assuming the spherical symmetry. 
For smoothing, we consider the cylindrical volume.
The width of the cylinder matches to the angular resolution~$\Delta \theta$ and the length corresponds to the spatial resolution in the line-of-sight direction determined by the frequency resolution~$\Delta \nu$.
To obtain the smoothed 21-cm signal $\delta T_{\mathrm b}$ which is defined in eq.~(\ref{eq:dtb2}) around the AGN, we smooth the signal with the cylindrical volume at whose center the AGN locates.

To discuss the detectability by SKA, we compare the signal $|\delta T_{\rm b}|$ with the noise level of the SKA observation.
The noise of an interferometer for an observation wavelength $\lambda_{\rm obs}$ is written in terms of the brightness temperature as \citep{2006PhR...433..181F}
\begin{eqnarray} 
  \delta T_{\rm{N}}(\lambda_{\rm obs}) & = &
    \frac{\lambda^{2}_{\rm obs}}{\Delta \theta^{2}A_{\rm{eff}}}\frac{T_{\rm{sys}}}{\sqrt{\Delta \nu t_{\rm{obs}}}} \nonumber \\
  & \simeq & 20\ [{\rm mK}]\left(\frac{A_{\rm eff}}{10^4\ [{\rm m^2]}}\right)^{-1} {\left(\frac{\Delta \theta}{10'}\right)}^{-2}{\left(\frac{1+z}{10}\right)}^4 \nonumber \\ 
  &\quad& \quad \times {\left(\frac{\Delta \nu}{1\ [{\rm MHz}]}\frac{t_{\rm int}}{100\ [{\rm hr}]}\right)}^{-1/2}, \label{eq:noise_eq}
\end{eqnarray}
where $\Delta \theta$ is the angular resolution, $\Delta \nu$ is the
frequency resolution,
$A_{\rm{eff}}$ is the effective collecting are a, $T_{\rm sys}$ is the
system temperature of the observation and $t_{\rm obs}$ is
the total observation time.
In order to obtain the second line of the equation, we use $\lambda_{\rm obs} = 21~{[\rm cm] } (1+z)$
For the SKA observation, we set $A_{\rm eff} = 10^4~[\rm m^2]$ and $t_{\rm
obs} = 1000$~[hours].
One of main contributions to $T_{\rm sys}$
is the synchrotron radiation
in the Milky Way. 
Therefore, for~$T_{\rm sys}$, we adopt the sky temperature at high
Galactic latitude,
\begin{eqnarray}
 T_{\rm sys} = 180~\left(\frac{\nu_{\rm obs}}{180~[{\rm MHz}]} \right)^{-2.6} \; \mbox{[K]} 
\end{eqnarray}
where $\nu_{\rm obs}$ is the frequency corresponding to
$\lambda_{\rm obs}$.

\begin{figure}
\begin{center}
\includegraphics[width=0.4\textwidth]{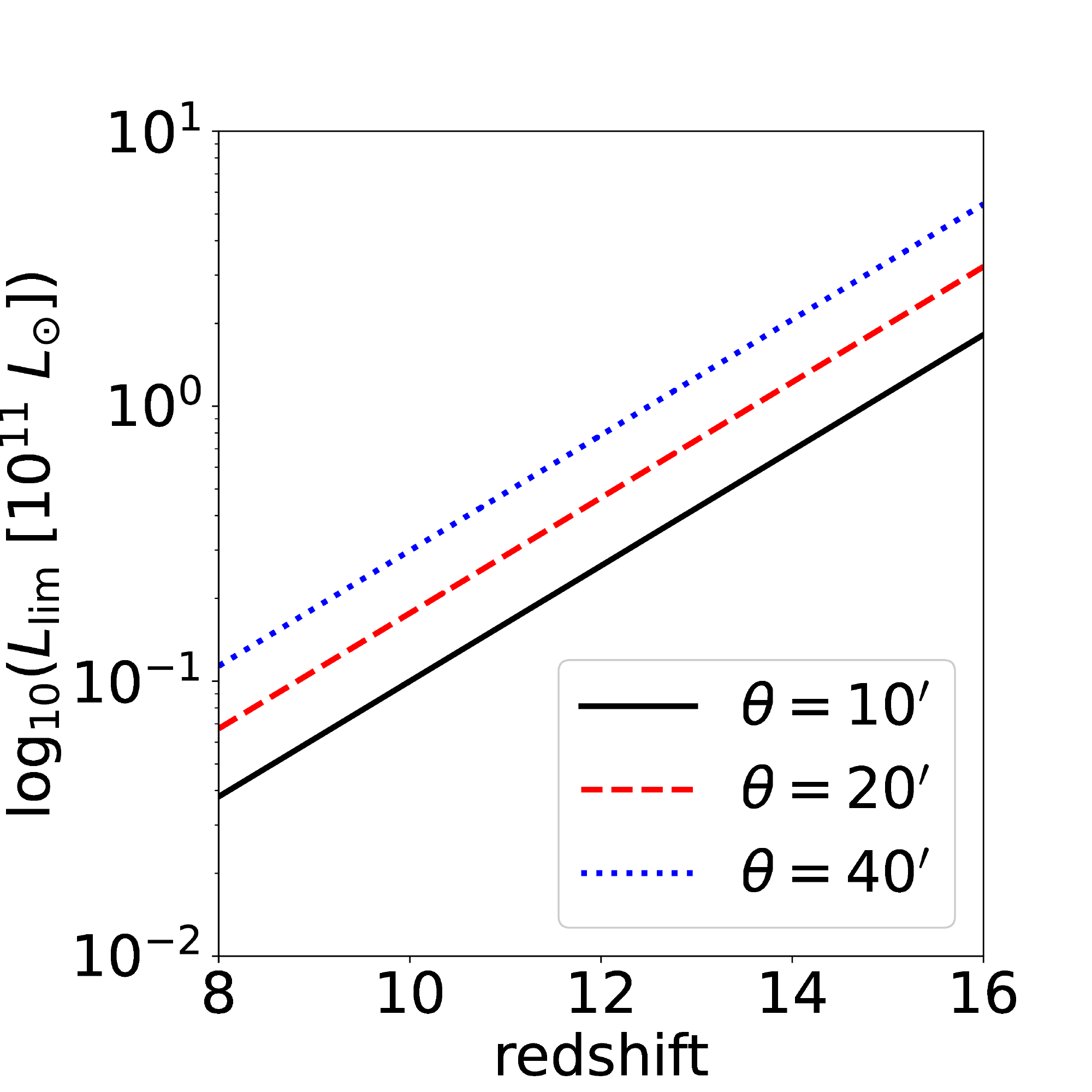}
\end{center}
\caption{The minimum luminosity for a set of $\Delta \theta$ as a function of redshift obtained from 2D convolution of $21\ {\rm cm}$ signal from AGN.}
\label{fig:minimum luminosity}
\end{figure}

We set our criterion for the detection to $|\delta T_{\rm b}|/
\delta T_{\rm{N}}>3$.
Since~$|\delta T_{\rm b}|$ becomes large
with increasing $L$ of the AGN, the detection criterion
can be represented 
as the minimum AGN luminosity
for the detection, $L_{\rm min}$.
Figure~\ref{fig:minimum luminosity} shows the redsfhit dependence
of~$L_{\rm min}$ for different angular resolutions,~$\Delta \theta$.
The noise has a strong redshift
dependence coming from 
\begin{eqnarray}
 \lambda_{\rm obs}^2 T_{\rm sys} \propto (1+z)^{4.6} \; .
\end{eqnarray}
Consequently, $L_{\rm min}$ monotonically grows in high redshifts. 
When the angular resolution becomes large,
the noise decreases according to Eq.~(\ref{eq:noise_eq}).
However the smoothing volume also increases.
Accordingly, the signal is diluted by the smoothing 
and a large luminosity is required for the detection with
a large angular resolution.

\subsection{Fisher matrix analysis}

Now we consider the galaxy number counts of AGNs expected to be detected by SKA.
Suppose that we perform the 21 cm observation 
in some redshift range.
We divide the redshift into some redshift bins
and count the number of detected AGNs in 
each redshift bin.
Besides, conducting the data analysis with different angular resolution~(or using different range of $k$-modes for the image reconstruction?), we also count the detected AGN number for each configuration of the angular resolution. 
Therefore, we can obtain
the number count in two dimensional bins for the redshift and angular resolution.

When we model the AGN LF,
we can evaluate the expected number counts of AGNs whose luminosities are larger than $L_{\rm min}$ 
$N_{i,j}$, from
\begin{eqnarray}
  N_{i,j} = 4 \pi f_{\rm sky} \int_{z_{i,\rm min}}^{z_{i,\rm max}} {\rm d}z \int_{L_{{\rm min},j}} \frac{R^2(z)}{H(z)} \frac{{\rm d}n}{{\rm d}L} {\rm d} L \; , \nonumber \\
\end{eqnarray} 
where $R(z)$ is the comoving distance to the redshift~$z$,
${z_{i,\rm min}} $ and $ {z_{i,\rm max}}$ are the minimum and maximum redshifts in the $i$-th redshift bin, $f_{\rm sky}$ is the sky fraction of the observation.
Since SKA will have $5\times 5$ square degree field of view, we set $f_{\rm sky}$ to be $f_{\rm sky} =0.0006$. 
In the equation, as shown in the previous section, $L_{\rm min}$ is a function of the redshift~$z$ and the angular resolution $\Delta \theta_j$ related to the $j$-th bin of the angular resolution.

The number count depends on the redshift and angular resolution.
Therefore, we conduct the Fisher matrix analysis for our three fiducial DP models, considering the number count for several redshift bins with different angular resolution set, $\{\Delta \theta_i \}$.
Fixing the angular resolution, we can probe the redshift evolution of the AGN LF, in particular, $\beta_1$ and $\beta_2$.
On the other hand, in a fixed redshift, the number counts on different angular resolution is expected to provide the luminosity dependence of the LF, $\alpha$ and $L_*$. 

We take the assumption that the likelihood distribution for the parameters is a multivariate Gaussian and takes the maximum value at the fiducial parameters.
Therefore, the Fisher matrix is given by
\begin{eqnarray}
F_{\mu \nu} = \sum_{i,j} \frac{1}{\sigma_{i,j}^2} \left(\frac{\partial N_{i,j}}{\partial \theta_\mu} 
\frac{\partial N_{i,j}}{\partial \theta_\nu} \right)
\end{eqnarray}
where $N_{i,j}$ is the number of the detected AGNs in the $i$-th bin of the observed redshifts with the $j$-th angular resolution.
In the equation,
$\sigma_{i,j}$ is the variance of $N_{i,j}$ and set to $\sigma_{i,j}^{-2} =N_{i,j}$, because
we assume that the detected number in the each bin follows the Poisson statistics.

\begin{figure*}
\begin{center}
\includegraphics[clip, width = \textwidth]{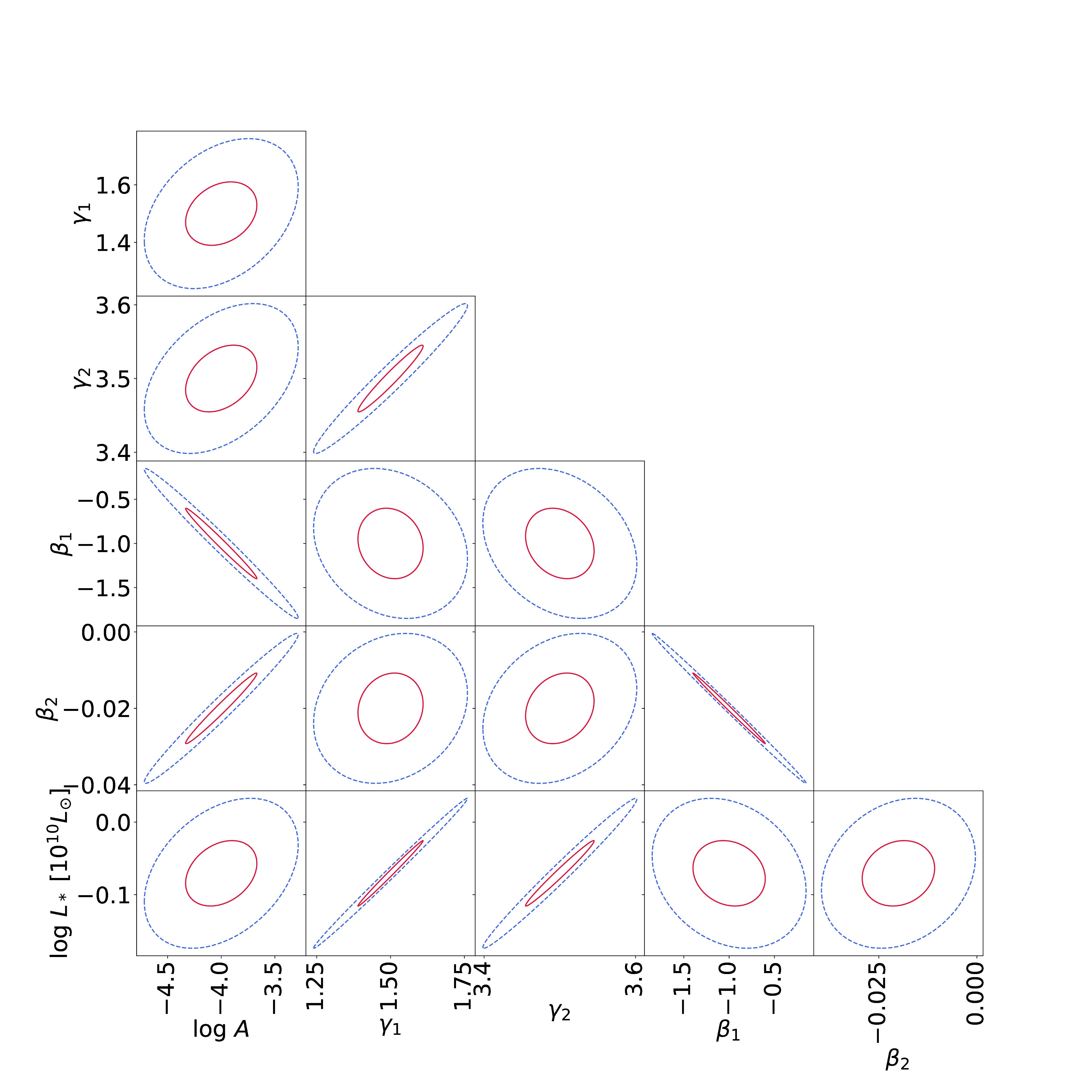}
\end{center}
\caption{The result of parameter constraint with Fisher analysis. The redshift and array configuration are divided into 10 bins for outer dashed elliptical and 20 bins for inner solid elliptical.}
\label{fig:fisherDP}
\end{figure*}

In this paper we take the ranges of $z$ in $[10,15]$ and $\Delta \theta$ in $[5,40] $ in the unit of the arc minutes. Dividing both $z$ and $\Delta \theta$ into 10~and 20~bins, we conduct the Fisher matrix analysis on our three fiducial models.
In Figure~\ref{fig:fisherDP}, we show the $1\mbox{-}\sigma$ error elliptical for 10~ (outside blue) and 20~ (inside red) bins for the LF parameter set for Model~I. As is shown in the Figure~\ref{fig:fisherDP}, the obtained errors are quite small. In particular, the errors of $\{\gamma_1,\ \beta_2,\ \log{L_*}\}$ are in a few percentage levels. On the other hand the errors of $\{A,\ \gamma_2,\ \beta_1\}$ are relatively large, compared with constrains on other parameters. However, as we can see later, SKA can determine these parameters in a factor level for Model~I

Figure~\ref{fig:fisherDP} also shows that there exist some strong correlations in the parameter sets. These correlation appears in order to compensate the increment~(decrements) due to a parameter by other parameters. For example, when we increase $\beta_1$, the AGN LF does not suppress on high redshifts. Therefore, to compensate it, small $A$ or large negative $\beta_2$ is preferred.

\begin{figure}
\begin{center}
  \includegraphics[width=0.5\textwidth]{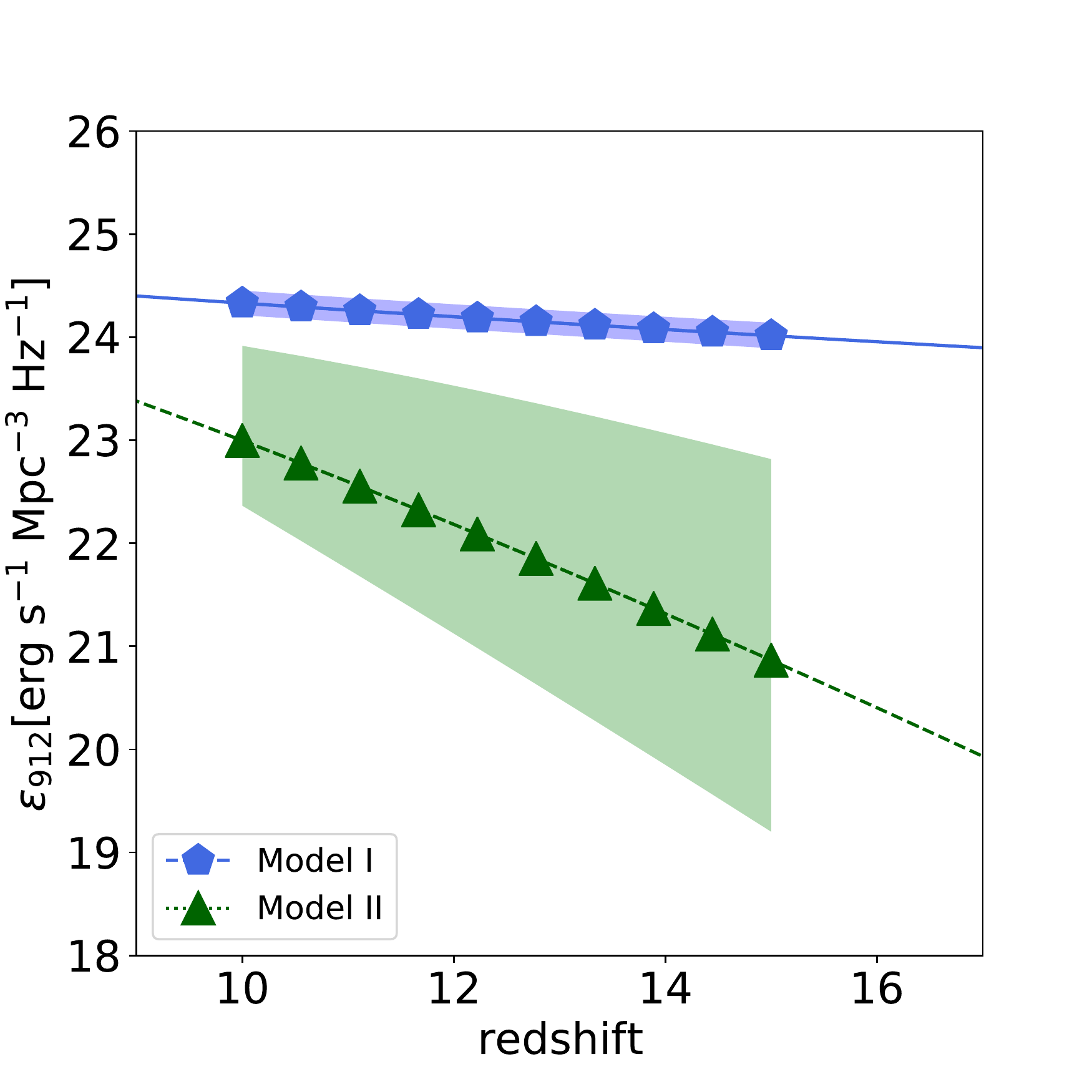} 
\end{center}
 \caption{The expected emissivity and estimated errors for $20\times 20$ configurations. The colored is derived by considering $1\mbox{-}\sigma$ error for the AGN LF parameters.}
\label{fig:e912_e}
\end{figure}

We summarize the $1\mbox{-}\sigma$ errors for all our fiducial models in Table~\ref{tab:tab4}. When we take the low LF model~(Model~III), the error becomes large since the number of observable galaxy significantly decreases. We found that the strong correlations shown in Figure~\ref{fig:fisherDP} also arises in the Model~II and III. 
Figure~\ref{fig:e912_e} tells us the impact of these errors on the determination of the LF in terms of the emissivity. Here the colored shaded regions represent $1\mbox{-}\sigma$ error for each fiducial model.
One can see that, when AGN luminosity distribution is given in the high LF model~(Model~I), SKA can reconstruct the LF. However, for Model 2, the reconstruction becomes worse, in particular, in high redshifts.

For simplicity, we assume that the AGN number count for each bin of the redshift and angular resolution is independent each other. Therefore, the errors of the LF parameters depends on the number of the bins. As we decrease the bin numbers, the error becomes large. For example, if we take 10 bins for the redshift and the angular resolution,
the error region increase two time larger than in Figure~\ref{fig:e912_e} and, as a result, we cannot determine the LF parameters even for the Model~II.

\begin{table*}
  \begin{center}
      \begin{tabular}{l|cccccc}
      \hline
      & $\delta\log \mathscr{A}$ &$\delta \gamma_1$&$\delta \gamma_2$&$\delta \beta_1$&$\delta \beta_2$& $\delta \log{L_*}$ \\ \hline
  Model~I  & $0.22$ & $7.3\times 10^{-2}$ & $3.0\times10^{-2}$ & $2.6\times10^{-1}$ & $6.1\times 10^{-3}$ & $0.030$\\ \hline
  Model~II & 2.3 & $4.5\times 10^{-1}$ & $1.8\times10^{-1}$ & 2.9 & $7.2\times 10^{-2}$ & $0.18$\\ \hline 
  Model~III  & $25$ & 3.1 & 1.1 & $3.1\times 10^1$ & 1.2 & $0.81$ \\ \hline
  \end{tabular}
  \end{center}
  \caption{1$\sigma$ errors for model parameters for ($20\times 20$)-bin configurations in our AGN model.}
  \label{tab:tab4}
\end{table*}

\begin{figure}
\begin{center}
  \includegraphics[width=0.5\textwidth]{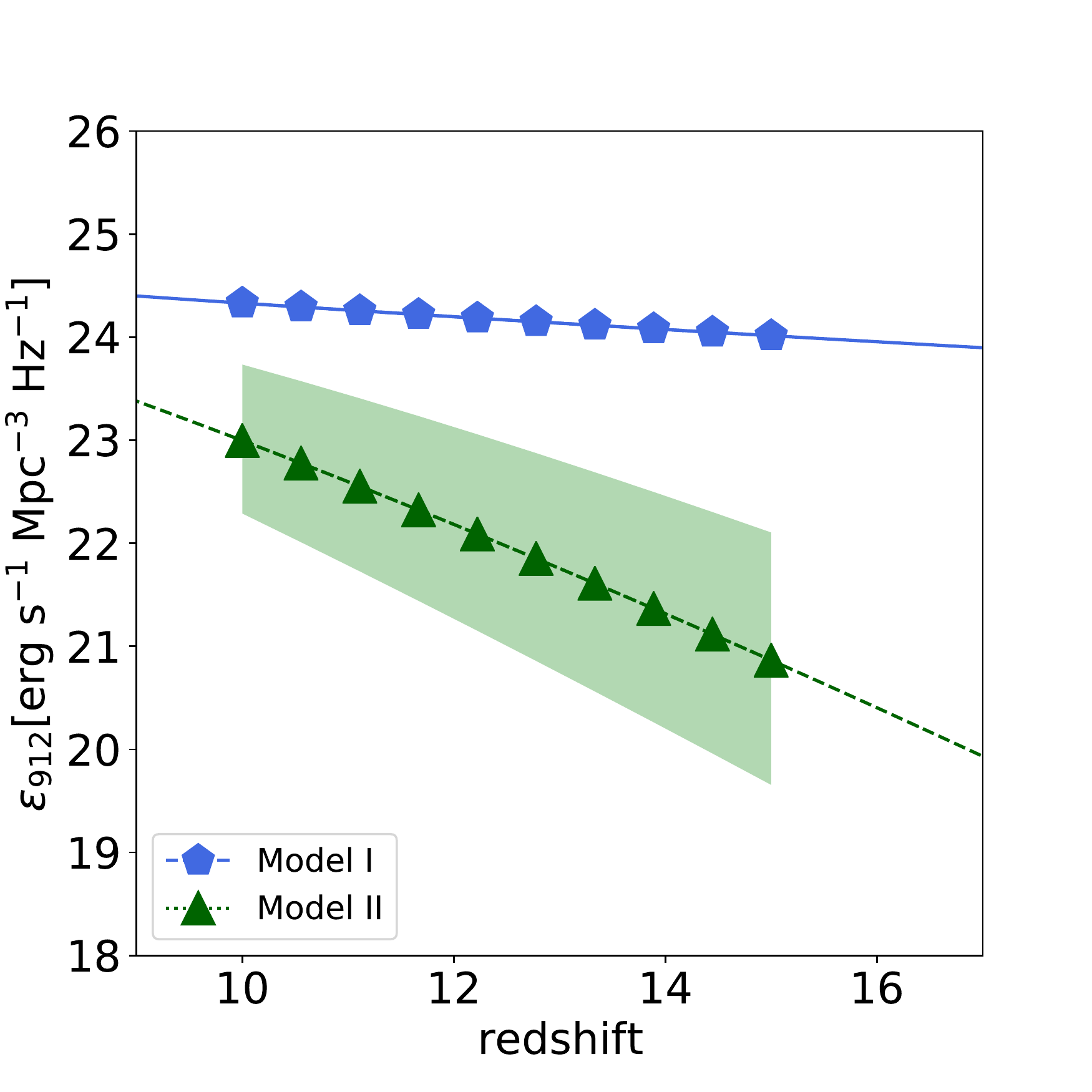} 
 \end{center}
\caption{Bins of galaxy number counts are divided into 20x20 observational configurations but flux limit is assumed to be $L_{\rm lim} = 1/10L_{\rm lim,0}$.}
\label{fig:e912_e2}
\end{figure}

At the last of this section,
we discuss the impact of the observation noise on the reconstruction of the AGN function.
In order to evaluate this impact,
we consider an observation which has 
sensitivity enough to detect 10 times lower luminosity AGNs than for SKA I.
We plot the result of the reconstruction 
in Figure~\ref{fig:e912_e2}.
For Model~II in high redshifts 
we can see the improvement for the reconstruction.
However,
since the SKA I sensitivity is enough to
measure the number counts for Model~I and Model~II in low redshifts, 
the reconstruction is not improved even for the better sensitivity.
Besides, it is impossible to obtain reasonable constraints on the parameters of Model~III even in the better 
sensitivity.

\section{Summary}
\label{sec:Summary}

AGNs can heat up the surrounding IGM gas by their UV and X-ray emission. 
Before the epoch of reionization, the heated neutral IGM can create the observable 21-cm emissions. Therefore, AGNs 
in high redshifts are promising targets in future redshifted 21-cm observations.
In this paper we have investigated the AGN number count by future 21-cm observation as a probe of the AGN LF in high redshifts.

First we have evaluated the 21-cm signals of AGNs before the epoch of reionization.
Assuming the power-law spectrum of the AGN UV and X-ray emission, we have calculated the radial profile of the IGM ionization fraction and temperature around an AGN by solving the radiative transfer of the AGN emission.
We have shown that AGNs can heat up the IGM gas even at the cosmological distance, i.e., comoving 10~Mpc, depending on the luminosity of AGNs.
As a result, AGNs can create the 21-cm signals 
whose angular size is roughly in the order of 10 arc-minutes.

The minimum amplitude of the 21 cm signal for the detection depends on the redshift and the angular resolution of observations. Therefore measuring the dependence of the AGN number count on the redshift and angular resolution allow us to probe the LF of AGNs.
To demonstrate the LF by AGN number count we propose the simple analytic form of the LF with several parameters which can recover well the emissivity of two typical empirical models, \citet{2015ApJ...813L...8M} and \citet{2007ApJ...654..731H}. 

Considering the 21-cm observations by SKA in the redshift range from $z=15$ to $z=10$ with the arc-minutes angular resolution,
we perform the Fisher matrix analysis to investigate how AGN number count can recover the LF.
We have found that SKA can probe the LF which is similar to the one suggested by \citet{2015ApJ...813L...8M}. 
However for the type propsed by \citet{2007ApJ...654..731H}, which is three order of magnitude smaller than the one by \citet{2015ApJ...813L...8M} in the emissivity,
it is difficult for SKA to determine the LF parameters.

\begin{ack}
This work has been supported by JSPS (Grant numbers: JPJSCCA20200002,17H01110, 18K03616, and 19H05076) and JST AIP Acceleration Research Grant Number JP20317829. 
This work has also been supported in part by the Sumitomo Foundation Fiscal
2018 Grant for Basic Science Research Projects (180923), and the Collaboration Funding of the Institute of Statistical Mathematics ``New Development of the Studies on Galaxy Evolution with a Method of Data Science''.
\end{ack}

\bibliography{SKA}

\begin{thebibliography}{}
\expandafter\ifx\csname natexlab\endcsname\relax\def\natexlab#1{#1}\fi

\bibitem[{{Alvarez} {et~al.}(2009){Alvarez}, {Wise}, \&
  {Abel}}]{2009ApJ...701L.133A}
{Alvarez}, M.~A., {Wise}, J.~H., \& {Abel}, T. 2009, \apj, 701, L133

\bibitem[{{Ba{\~n}ados} {et~al.}(2018){Ba{\~n}ados}, {Connor}, {Stern},
  {Mulchaey}, {Fan}, {Decarli}, {Farina}, {Mazzucchelli}, {Venemans}, {Walter},
  {Wang}, \& {Yang}}]{2018ApJ...856L..25B}
{Ba{\~n}ados}, E., {Connor}, T., {Stern}, D., {et~al.} 2018, \apj, 856, L25

\bibitem[{{Bean} \& {Magueijo}(2002)}]{2002PhRvD..66f3505B}
{Bean}, R., \& {Magueijo}, J. 2002, \prd, 66, 063505

\bibitem[{{Begelman} {et~al.}(2006){Begelman}, {Volonteri}, \&
  {Rees}}]{2006MNRAS.370..289B}
{Begelman}, M.~C., {Volonteri}, M., \& {Rees}, M.~J. 2006, \mnras, 370, 289

\bibitem[{{Bongiorno} {et~al.}(2007){Bongiorno}, {Zamorani}, {Gavignaud},
  {Marano}, {Paltani}, {Mathez}, {M{\o}ller}, {Picat}, {Cirasuolo},
  {Lamareille}, {Bottini}, {Garilli}, {Le Brun}, {Le F{\`e}vre}, {Maccagni},
  {Scaramella}, {Scodeggio}, {Tresse}, {Vettolani}, {Zanichelli}, {Adami},
  {Arnouts}, {Bardelli}, {Bolzonella}, {Cappi}, {Charlot}, {Ciliegi},
  {Contini}, {Foucaud}, {Franzetti}, {Guzzo}, {Ilbert}, {Iovino}, {McCracken},
  {Marinoni}, {Mazure}, {Meneux}, {Merighi}, {Pell{\`o}}, {Pollo}, {Pozzetti},
  {Radovich}, {Zucca}, {Hatziminaoglou}, {Polletta}, {Bondi}, {Brinchmann},
  {Cucciati}, {de la Torre}, {Gregorini}, {Mellier}, {Merluzzi}, {Temporin},
  {Vergani}, \& {Walcher}}]{2007A&A...472..443B}
{Bongiorno}, A., {Zamorani}, G., {Gavignaud}, I., {et~al.} 2007, \aap, 472, 443

\bibitem[{{Bouwens} {et~al.}(2012){Bouwens}, {Illingworth}, {Oesch}, {Trenti},
  {Labb{\'e}}, {Franx}, {Stiavelli}, {Carollo}, {van Dokkum}, \&
  {Magee}}]{2012ApJ...752L...5B}
{Bouwens}, R.~J., {Illingworth}, G.~D., {Oesch}, P.~A., {et~al.} 2012, \apjl,
  752, L5

\bibitem[{{Brandt} \& {Alexander}(2015)}]{2015A&ARv..23....1B}
{Brandt}, W.~N., \& {Alexander}, D.~M. 2015, \aapr, 23, 1

\bibitem[{{Cattaneo} {et~al.}(2009){Cattaneo}, {Faber}, {Binney}, {Dekel},
  {Kormendy}, {Mushotzky}, {Babul}, {Best}, {Br{\"u}ggen}, {Fabian}, {Frenk},
  {Khalatyan}, {Netzer}, {Mahdavi}, {Silk}, {Steinmetz}, \&
  {Wisotzki}}]{2009Natur.460..213C}
{Cattaneo}, A., {Faber}, S.~M., {Binney}, J., {et~al.} 2009, \nat, 460, 213

\bibitem[{{Chen} \& {Miralda-Escud{\'e}}(2008)}]{2008ApJ...684...18C}
{Chen}, X., \& {Miralda-Escud{\'e}}, J. 2008, \apj, 684, 18

\bibitem[{{Christian} \& {Loeb}(2013)}]{2013JCAP...09..014C}
{Christian}, P., \& {Loeb}, A. 2013, Journal of Cosmology and Astroparticle
  Physics, 2013, 014

\bibitem[{{Ciardi} \& {Madau}(2003)}]{2003ApJ...596....1C}
{Ciardi}, B., \& {Madau}, P. 2003, \apj, 596, 1

\bibitem[{{David} {et~al.}(1992){David}, {Jones}, \&
  {Forman}}]{1992ApJ...388...82D}
{David}, L.~P., {Jones}, C., \& {Forman}, W. 1992, \apj, 388, 82

\bibitem[{{Dijkstra} {et~al.}(2004){Dijkstra}, {Haiman}, \&
  {Loeb}}]{2004ApJ...613..646D}
{Dijkstra}, M., {Haiman}, Z., \& {Loeb}, A. 2004, \apj, 613, 646

\bibitem[{{D{\"u}chting}(2004)}]{2004PhRvD..70f4015D}
{D{\"u}chting}, N. 2004, \prd, 70, 064015

\bibitem[{{Eisenstein} \& {Loeb}(1995)}]{1995ApJ...443...11E}
{Eisenstein}, D.~J., \& {Loeb}, A. 1995, \apj, 443, 11

\bibitem[{{Fan} {et~al.}(2006){Fan}, {Strauss}, {Richards}, {Hennawi},
  {Becker}, {White}, {Diamond-Stanic}, {Donley}, {Jiang}, {Kim}, {Vestergaard},
  {Young}, {Gunn}, {Lupton}, {Knapp}, {Schneider}, {Brandt}, {Bahcall},
  {Barentine}, {Brinkmann}, {Brewington}, {Fukugita}, {Harvanek}, {Kleinman},
  {Krzesinski}, {Long}, {Neilsen}, {Nitta}, {Snedden}, \&
  {Voges}}]{2006AJ....131.1203F}
{Fan}, X., {Strauss}, M.~A., {Richards}, G.~T., {et~al.} 2006, \aj, 131, 1203

\bibitem[{{Ferrarese} \& {Ford}(2005)}]{2005SSRv..116..523F}
{Ferrarese}, L., \& {Ford}, H. 2005, \ssr, 116, 523

\bibitem[{{Ferrarese} \& {Merritt}(2000)}]{2000ApJ...539L...9F}
{Ferrarese}, L., \& {Merritt}, D. 2000, \apj, 539, L9

\bibitem[{{Fialkov} {et~al.}(2014){Fialkov}, {Barkana}, \&
  {Visbal}}]{2014Natur.506..197F}
{Fialkov}, A., {Barkana}, R., \& {Visbal}, E. 2014, \nat, 506, 197

\bibitem[{{Field}(1958)}]{1958PIRE...46..240F}
{Field}, G.~B. 1958, Proceedings of the IRE, 46, 240

\bibitem[{{Fontanot} {et~al.}(2012){Fontanot}, {Cristiani}, \&
  {Vanzella}}]{2012MNRAS.425.1413F}
{Fontanot}, F., {Cristiani}, S., \& {Vanzella}, E. 2012, \mnras, 425, 1413

\bibitem[{{Fukugita} \& {Kawasaki}(1994)}]{1994MNRAS.269..563F}
{Fukugita}, M., \& {Kawasaki}, M. 1994, \mnras, 269, 563

\bibitem[{{Furlanetto} {et~al.}(2006){Furlanetto}, {Oh}, \&
  {Briggs}}]{2006PhR...433..181F}
{Furlanetto}, S.~R., {Oh}, S.~P., \& {Briggs}, F.~H. 2006, \physrep, 433, 181

\bibitem[{{Gebhardt} {et~al.}(2000){Gebhardt}, {Bender}, {Bower}, {Dressler},
  {Faber}, {Filippenko}, {Green}, {Grillmair}, {Ho}, {Kormendy}, {Lauer},
  {Magorrian}, {Pinkney}, {Richstone}, \& {Tremaine}}]{2000ApJ...539L..13G}
{Gebhardt}, K., {Bender}, R., {Bower}, G., {et~al.} 2000, \apj, 539, L13

\bibitem[{{Giallongo} {et~al.}(2015){Giallongo}, {Grazian}, {Fiore}, {Fontana},
  {Pentericci}, {Vanzella}, {Dickinson}, {Kocevski}, {Castellano}, {Cristiani},
  {Ferguson}, {Finkelstein}, {Grogin}, {Hathi}, {Koekemoer}, {Newman}, \&
  {Salvato}}]{2015A&A...578A..83G}
{Giallongo}, E., {Grazian}, A., {Fiore}, F., {et~al.} 2015, \aap, 578, A83

\bibitem[{{Glikman} {et~al.}(2011){Glikman}, {Djorgovski}, {Stern}, {Dey},
  {Jannuzi}, \& {Lee}}]{2011ApJ...728L..26G}
{Glikman}, E., {Djorgovski}, S.~G., {Stern}, D., {et~al.} 2011, \apj, 728, L26

\bibitem[{{Hopkins} {et~al.}(2007){Hopkins}, {Richards}, \&
  {Hernquist}}]{2007ApJ...654..731H}
{Hopkins}, P.~F., {Richards}, G.~T., \& {Hernquist}, L. 2007, \apj, 654, 731

\bibitem[{{Koopmans} {et~al.}(2015){Koopmans}, {Pritchard}, {Mellema},
  {Aguirre}, {Ahn}, {Barkana}, {van Bemmel}, {Bernardi}, {Bonaldi}, {Briggs},
  {de Bruyn}, {Chang}, {Chapman}, {Chen}, {Ciardi}, {Dayal}, {Ferrara},
  {Fialkov}, {Fiore}, {Ichiki}, {Illiev}, {Inoue}, {Jelic}, {Jones}, {Lazio},
  {Maio}, {Majumdar}, {Mack}, {Mesinger}, {Morales}, {Parsons}, {Pen},
  {Santos}, {Schneider}, {Semelin}, {de Souza}, {Subrahmanyan}, {Takeuchi},
  {Vedantham}, {Wagg}, {Webster}, {Wyithe}, {Datta}, \&
  {Trott}}]{2015aska.confE...1K}
{Koopmans}, L., {Pritchard}, J., {Mellema}, G., {et~al.} 2015, in Advancing
  Astrophysics with the Square Kilometre Array (AASKA14), 1

\bibitem[{{Kuhlen} \& {Madau}(2005)}]{2005MNRAS.363.1069K}
{Kuhlen}, M., \& {Madau}, P. 2005, \mnras, 363, 1069

\bibitem[{{Kuhlen} {et~al.}(2006){Kuhlen}, {Madau}, \&
  {Montgomery}}]{2006ApJ...637L...1K}
{Kuhlen}, M., {Madau}, P., \& {Montgomery}, R. 2006, \apj, 637, L1

\bibitem[{{Kulkarni} {et~al.}(2017){Kulkarni}, {Choudhury}, {Puchwein}, \&
  {Haehnelt}}]{2017MNRAS.469.4283K}
{Kulkarni}, G., {Choudhury}, T.~R., {Puchwein}, E., \& {Haehnelt}, M.~G. 2017,
  \mnras, 469, 4283

\bibitem[{{Lawrence} {et~al.}(1999){Lawrence}, {Rowan-Robinson}, {Ellis},
  {Frenk}, {Efstathiou}, {Kaiser}, {Saunders}, {Parry}, {Xiaoyang}, \&
  {Crawford}}]{1999MNRAS.308..897L}
{Lawrence}, A., {Rowan-Robinson}, M., {Ellis}, R.~S., {et~al.} 1999, \mnras,
  308, 897

\bibitem[{{Loeb} \& {Rasio}(1994)}]{1994ApJ...432...52L}
{Loeb}, A., \& {Rasio}, F.~A. 1994, \apj, 432, 52

\bibitem[{{Lonsdale} {et~al.}(1990){Lonsdale}, {Hacking}, {Conrow}, \&
  {Rowan-Robinson}}]{1990ApJ...358...60L}
{Lonsdale}, C.~J., {Hacking}, P.~B., {Conrow}, T.~P., \& {Rowan-Robinson}, M.
  1990, \apj, 358, 60

\bibitem[{{Madau} \& {Haardt}(2015)}]{2015ApJ...813L...8M}
{Madau}, P., \& {Haardt}, F. 2015, \apj, 813, L8

\bibitem[{{Madau} {et~al.}(1997){Madau}, {Meiksin}, \&
  {Rees}}]{1997ApJ...475..429M}
{Madau}, P., {Meiksin}, A., \& {Rees}, M.~J. 1997, \apj, 475, 429

\bibitem[{{Madau} \& {Rees}(2001)}]{2001ApJ...551L..27M}
{Madau}, P., \& {Rees}, M.~J. 2001, \apj, 551, L27

\bibitem[{{Magorrian} {et~al.}(1998){Magorrian}, {Tremaine}, {Richstone},
  {Bender}, {Bower}, {Dressler}, {Faber}, {Gebhardt}, {Green}, {Grillmair},
  {Kormendy}, \& {Lauer}}]{1998AJ....115.2285M}
{Magorrian}, J., {Tremaine}, S., {Richstone}, D., {et~al.} 1998, \aj, 115, 2285

\bibitem[{{Marconi} \& {Hunt}(2003)}]{2003ApJ...589L..21M}
{Marconi}, A., \& {Hunt}, L.~K. 2003, \apj, 589, L21

\bibitem[{{Masters} {et~al.}(2012){Masters}, {Capak}, {Salvato}, {Civano},
  {Mobasher}, {Siana}, {Hasinger}, {Impey}, {Nagao}, {Trump}, {Ikeda}, {Elvis},
  \& {Scoville}}]{2012ApJ...755..169M}
{Masters}, D., {Capak}, P., {Salvato}, M., {et~al.} 2012, \apj, 755, 169

\bibitem[{{Mauch} \& {Sadler}(2007)}]{2007MNRAS.375..931M}
{Mauch}, T., \& {Sadler}, E.~M. 2007, \mnras, 375, 931

\bibitem[{{Mitra} {et~al.}(2018){Mitra}, {Choudhury}, \&
  {Ferrara}}]{2018MNRAS.473.1416M}
{Mitra}, S., {Choudhury}, T.~R., \& {Ferrara}, A. 2018, \mnras, 473, 1416

\bibitem[{{Mortlock} {et~al.}(2011){Mortlock}, {Warren}, {Venemans}, {Patel},
  {Hewett}, {McMahon}, {Simpson}, {Theuns}, {Gonz{\'a}les-Solares}, {Adamson},
  {Dye}, {Hambly}, {Hirst}, {Irwin}, {Kuiper}, {Lawrence}, \&
  {R{\"o}ttgering}}]{2011Natur.474..616M}
{Mortlock}, D.~J., {Warren}, S.~J., {Venemans}, B.~P., {et~al.} 2011, \nat,
  474, 616

\bibitem[{{Nanni} {et~al.}(2018){Nanni}, {Gilli}, {Vignali}, {Mignoli},
  {Comastri}, {Vanzella}, {Zamorani}, {Calura}, {Lanzuisi}, {Brusa}, {Tozzi},
  {Iwasawa}, {Cappi}, {Vito}, {Balmaverde}, {Costa}, {Risaliti}, {Paolillo},
  {Prandoni}, {Liuzzo}, {Rosati}, {Chiaberge}, {Caminha}, {Sani}, {Cappelluti},
  \& {Norman}}]{2018A&A...614A.121N}
{Nanni}, R., {Gilli}, R., {Vignali}, C., {et~al.} 2018, \aap, 614, A121

\bibitem[{{Palanque-Delabrouille} {et~al.}(2013){Palanque-Delabrouille},
  {Magneville}, {Y{\`e}che}, {Eftekharzadeh}, {Myers}, {Petitjean},
  {P{\^a}ris}, {Aubourg}, {McGreer}, {Fan}, {Dey}, {Schlegel}, {Bailey},
  {Bizayev}, {Bolton}, {Dawson}, {Ebelke}, {Ge}, {Malanushenko},
  {Malanushenko}, {Oravetz}, {Pan}, {Ross}, {Schneider}, {Sheldon}, {Simmons},
  {Tinker}, {White}, \& {Willmer}}]{2013A&A...551A..29P}
{Palanque-Delabrouille}, N., {Magneville}, C., {Y{\`e}che}, C., {et~al.} 2013,
  \aap, 551, A29

\bibitem[{{Parsa} {et~al.}(2018){Parsa}, {Dunlop}, \&
  {McLure}}]{2018MNRAS.474.2904P}
{Parsa}, S., {Dunlop}, J.~S., \& {McLure}, R.~J. 2018, \mnras, 474, 2904

\bibitem[{{Pons} {et~al.}(2020){Pons}, {McMahon}, {Banerji}, \&
  {Reed}}]{2020MNRAS.491.3884P}
{Pons}, E., {McMahon}, R.~G., {Banerji}, M., \& {Reed}, S.~L. 2020, \mnras,
  491, 3884

\bibitem[{{Pritchard} \& {Furlanetto}(2007)}]{2007MNRAS.376.1680P}
{Pritchard}, J.~R., \& {Furlanetto}, S.~R. 2007, \mnras, 376, 1680

\bibitem[{{Saunders} {et~al.}(1990){Saunders}, {Rowan-Robinson}, {Lawrence},
  {Efstathiou}, {Kaiser}, {Ellis}, \& {Frenk}}]{1990MNRAS.242..318S}
{Saunders}, W., {Rowan-Robinson}, M., {Lawrence}, A., {et~al.} 1990, \mnras,
  242, 318

\bibitem[{{Schulze} {et~al.}(2009){Schulze}, {Wisotzki}, \&
  {Husemann}}]{2009A&A...507..781S}
{Schulze}, A., {Wisotzki}, L., \& {Husemann}, B. 2009, \aap, 507, 781

\bibitem[{{Shull} \& {van Steenberg}(1985)}]{1985ApJ...298..268S}
{Shull}, J.~M., \& {van Steenberg}, M.~E. 1985, \apj, 298, 268

\bibitem[{{Tashiro} \& {Sugiyama}(2013)}]{2013MNRAS.435.3001T}
{Tashiro}, H., \& {Sugiyama}, N. 2013, \mnras, 435, 3001

\bibitem[{{Ueda} {et~al.}(2014){Ueda}, {Akiyama}, {Hasinger}, {Miyaji}, \&
  {Watson}}]{2014ApJ...786..104U}
{Ueda}, Y., {Akiyama}, M., {Hasinger}, G., {Miyaji}, T., \& {Watson}, M.~G.
  2014, \apj, 786, 104

\bibitem[{{Whalen} \& {Fryer}(2012)}]{2012ApJ...756L..19W}
{Whalen}, D.~J., \& {Fryer}, C.~L. 2012, \apj, 756, L19

\bibitem[{{Wouthuysen}(1952)}]{1952AJ.....57R..31W}
{Wouthuysen}, S.~A. 1952, \aj, 57, 31

\bibitem[{{Wu} {et~al.}(2015){Wu}, {Wang}, {Fan}, {Yi}, {Zuo}, {Bian}, {Jiang},
  {McGreer}, {Wang}, {Yang}, {Yang}, {Thompson}, \&
  {Beletsky}}]{2015Natur.518..512W}
{Wu}, X.-B., {Wang}, F., {Fan}, X., {et~al.} 2015, \nat, 518, 512

\bibitem[{{Yajima} \& {Li}(2014)}]{2014MNRAS.445.3674Y}
{Yajima}, H., \& {Li}, Y. 2014, \mnras, 445, 3674

\bibitem[{{Yoshiura} {et~al.}(2017){Yoshiura}, {Hasegawa}, {Ichiki}, {Tashiro},
  {Shimabukuro}, \& {Takahashi}}]{2017MNRAS.471.3713Y}
{Yoshiura}, S., {Hasegawa}, K., {Ichiki}, K., {et~al.} 2017, \mnras, 471, 3713

\bibitem[{{Zaroubi} \& {Silk}(2005)}]{2005MNRAS.360L..64Z}
{Zaroubi}, S., \& {Silk}, J. 2005, \mnras, 360, L64

\bibitem[{{Zaroubi} {et~al.}(2007){Zaroubi}, {Thomas}, {Sugiyama}, \&
  {Silk}}]{2007MNRAS.375.1269Z}
{Zaroubi}, S., {Thomas}, R.~M., {Sugiyama}, N., \& {Silk}, J. 2007, \mnras,
  375, 1269

\end{thebibliography}
\bibliographystyle{apj}

\end{document}